\documentclass[%
 reprint,
 amsmath,amssymb,
 aps,
]{revtex4-2}

\usepackage{graphicx}
\usepackage{dcolumn}
\usepackage{bm}

\begin{document}

\preprint{APS/123-QED}

\title{Fast forward-and-backward creation of non-equilibrium Gibbs state in driven dissipative quantum systems}

\author{Katsuhiro Nakamura}
\email{Corresponding author: ulbkatsu60@gmail.com}
\affiliation{Faculty of Physics, National University of Uzbekistan, Vuzgorodok, Tashkent 100174, Uzbekistan}
\author{Otabek Normukhamedov} 
\affiliation{Faculty of Physics, National University of Uzbekistan, Vuzgorodok, Tashkent 100174, Uzbekistan}
\author{Iwan Setiawan}
\affiliation{Department of Physics Education, University of Bengkulu, Kandang
Limun, Bengkulu 38371, Indonesia}
\author{Jasur Matrasulov}
\affiliation{Department of Electronics and Radio Engineering, Tashkent University of
Information Technologies, Amir Temur str., Tashkent 100084, Uzbekistan}

\date{\today}

\begin{abstract}
We present the fast-forward (FF) approach to
adiabatic time evolution of an open dissipative quantum system within the framework of the Gorini-Kosaakowski-Sudarshan-Lindblad (GKSL) equation. Taking a harmonic oscillator (HO) and a quantum spin as prototype systems under a rapid change of the driving field and in contact with the heat reservoir, we reveal the protocol which creates the non-equilibrium Gibbs state. After preparing the Gibbs state under the fixed external field in advance, we begin to change adiabatically the field. To keep a quasi-adiabatic version of the Gibbs state as the solution of GKSL equation, we solve the inverse problem to obtain the non-adiabatic corrections both to Hamiltonian and to the dissipator.  Then we apply the fast forward-and-backward (abbreviated as FF for convenience) time-rescaling, obtaining the FF variants of both Hamiltonian and detailed balance condition (DBC) among dissipative coefficients. The present scheme generates the time-periodic Gibbs state as if a limit-cycle solution of the innovated GKSL equation under the time-periodic driving field. 
Thanks to the FF variant of DBC, we see the energy conservation of
power and heat current of the system, and reversibility of whole system including a reservoir.
The requirement of positivity of the FF dissipative coefficients imposes a constraint on the speed of the FF driving.  At low temperature, the upper speed limit  of the FF periodic driving is found to be proportional to $\tilde{\beta}^{-1/2}\exp(a\tilde{\beta})$ with $\tilde{\beta}$ the scaled inverse temperature and $a$ the positive constant of $O(1)$.
\end{abstract}

\maketitle


\section{Introduction}
A shorter time in manufacturing products is becoming an important factor in nanotechnology.  
If we try to fabricate massive amount of nanoscale structures, we should accelerate the dynamics of each electron, atom or molecules to reach their desired target states in shorter time. In designing quantum computers, the coherence of systems is degraded by their interaction with the environment, and therefore the speedup of adiabatic quantum dynamics is highly desirable. 

Some time ago, Masuda and Nakamura \cite{MN-0} proposed a way to realize the fast-forwarding of arbitrary time evolution of  quantum systems by application of suitable driving forces. This kind of acceleration is called the fast forward (FF), which means to reproduce a series of events or a history of matters on a shortened time scale, like a rapid projection of movie films on the screen.  While the FF theory was originally
designed just for speed control of non-adiabatic dynamics, it was also developed towards the FF of adiabatic dynamics \cite{MN-1,MN-2}.
The FF scheme applied to the adiabatic quantum dynamics employs a trick of the infinitely-large time-scaling factor that compensates the adiabaticity, i.e., a very small growth rate of the adiabatic parameter. 
This theory provided a simple protocol to accelerate the quantum tunneling \cite{KN-1,KN-2} and adiabatic entanglement dynamics \cite{SG-1,SG-2}.  The FF theory is also applicable to dynamical construction of classical adiabatic invariant \cite{Jal-1} and semiclassical analysis \cite{Jal-2}. The FF theory constitutes an alternative way of shortcuts to adiabaticity (STA) \cite{STA-0,STA-1,STA-rev} devoted to tailor excitations in non-adiabatic processes.
The relationship between the schemes of FF theory and STA is explained in the review article \cite{MN-rev}. 

The FF theory, however, has not been cultivated on systems with decoherence and dissipation, except for a classical system \cite{NMI}.  
Recent progress in quantum information processing and quantum computing \cite{Niel,Wisen} requires precise monitoring of decoherence and dissipation in mesoscopic systems, and the Gorini-Kosaakowski-Sudarshan-Lindblad (GKSL) equation \cite{GKS,Lind,Alick} has received a growing attention as  a prototype of the open quantum master equation or quantum dynamical semigroup. The GKSL equation incorporates effectively the role of the environment in the form a completely-positive and trace-preserving (CPTP) map acting on the Markovian systems, and is known useful in areas in quantum logic gates and quantum thermodynamics. 

The present paper is concerned with application of the FF scheme to GKSL equation. Throughout the text, FF here means the abbreviation of fast forward-and-backward.
After the relaxation time, the density matrix of a system in a time-independent potential becomes to have a solution of the Gibbs state, i.e., a key concept of the equilibrium thermodynamics and statistical mechanics. 
If the potential will then begin to be time-periodic, the fate of the Gibbs state is not clear at all. The FF scheme, however,  lets the density matrix have a 
limit-cycle-like solution characterized with the time-periodic Gibbs state. 
We shall analytically explore this issue in prototype examples.

Sections from \ref{AOQD} to \ref{FF rescaling} are concerned with the FF of a driven dissipative quantum harmonic oscillator (HO).
We shall begin with HO in the constant confinement and its Gibbs state which is a fixed-point solution of the GKSL equation with dissipators whose coefficients satisfy the standard detailed balance condition (DBC) or KMS condition\cite{Kubo,Martin,Yablon}. Then the strategy of FF consists of 2 steps.  Firstly, assuming an extremely slow temporal change of the confining parameter, we make a quasi-adiabatic version of the Gibbs state satisfy the GKSL equation by adding the non-adiabatic corrections to Hamiltonian and to the dissipator. This is an inverse problem to be explained in Sections \ref{AOQD} and \ref{NAHD}. Secondly, to realize the adiabatic time evolution on shortened time scale,  we make FF time-rescaling in the quasi-adiabatic GKSL equation, predicting  the FF Hamiltonian and FF version of the DBC, 
to be described in Section \ref{FF rescaling}. Regarding the FF time as a period of periodic driving, the FF scheme is shown to create the time-periodic Gibbs state. To show a wide applicability of the FF scheme, we study another prototype system, i.e., a driven dissipative quantum spin system in Section \ref{AOQD2}. 
In Section \ref{implement},  to see the role of the FF variant of DBC we  investigate thermodynamic properties of  the nonequilibrium Gibbs state during its FF time evolution. 
This Section also describes a constraint on driving speed of the FF protocol and reveals the temperature dependence of the speed limit.
Summary and discussion are given in Section \ref{conclude}.
Appendix \ref{apdB} shows the equivalence and difference of dissipators between STA and FF approaches.

\section{Scheme of adiabatic open dissipative quantum dynamics}\label{AOQD}
We present the fast forward (FF) scheme of an open dissipative quantum dynamics under the time-dependent confinement, by using the GKSL's quantum master equation for the system's density operator \cite{GKS,Lind,Alick}.
Here the system consists of a particle with mass $m$ confined by a harmonic trap with frequency $\omega$ and has contact with heat reservoirs.  The present study is parallel to the FF approach to the classical stochastic heat engine developed by Nakamura, Matrasulov and Izumida  \cite{NMI}. 
The GKSL equation for the system's density operator $\hat{\rho}$ takes the form:
\begin{equation}\label{GKSL-orig}
\frac{\partial\hat{\rho}}{\partial t}=-\frac{i}{\hbar}[\hat{H_0},\hat{\rho}]+\hat{D}_0[\hat{\rho}]. 
\end{equation}
$\hat{H_0}$ is a scale-invariant Hamiltonian,
\begin{eqnarray}\label{hamil-orig}
\hat{H_0}(R)&=&\frac{\hat{p}^2}{2m}+\frac{1}{R^2}V_0\Bigl(\frac{\hat{x}}{R}\Bigr)\nonumber\\
    &=& -\frac{\hbar^2}{2m}\frac{\partial^2}{\partial x^2}+\frac{1}{2}m\omega^2(R)x^2,
\end{eqnarray}
where the adiabatic parameter $R$ denotes the confining length, 
\begin{equation}\label{width}
R=\sqrt{\frac{\hbar}{m\omega}},
\end{equation}
representing the width of the ground state wavefunction.
The instantaneous (i.e., adiabatic) eigenvalues and eigenstates are respectively given below:
\begin{equation}\label{eigenenerg}
    E_n(R)=\Bigl(n+\frac{1}{2}\Bigr)\hbar\omega(R), \;\;\;\;n=0,1,2,\ldots,
\end{equation}
\begin{equation}\label{eigen state}
    \langle x|n\rangle=\phi_n(x,R)=C_n \alpha^{\frac{1}{2}}e^{-\frac{\alpha^2}{2}x^2}H_n(\alpha x),
\end{equation}
where $H_n(\cdots)$ is the Hermite polynomial with $\alpha=\Bigl(\frac{m\omega(R)}{\hbar}\Bigr)^{\frac{1}{2}}$ and $C_n=(2^n n!\sqrt{\pi})^{-\frac{1}{2}}$.
The Hamiltonian in Eq.(\ref{hamil-orig}) can be quantized as 
\begin{equation}\label{diagonal}
    \hat{H_0}(R)=\Bigl(\hat{a}^+\hat{a}+\frac{1}{2}\Bigr)\hbar\omega(R),
\end{equation}
where $\hat{a}=\frac{1}{\sqrt{2\hbar}}\Bigl(\sqrt{m\omega}\hat{x}+i\frac{1}{\sqrt{m\omega}}\hat{p}\Bigr)$ and $\hat{a}^+=\frac{1}{\sqrt{2\hbar}}\Bigl(\sqrt{m\omega}\hat{x}-i\frac{1}{\sqrt{m\omega}}\hat{p}\Bigr)$ are energy lowering and raising operators, respectively, working on $\{\phi_n(x,R)\}$.
In Eq.(\ref{GKSL-orig}), $\hat{D}_0[\hat{\rho}]$ is the Lindblad dissipator defined by\cite{Lind} 
\begin{equation} \label{updown dissip}  
\hat{D}_0[\hat{\rho}]\equiv\sum_{\alpha=\uparrow,\downarrow}\hat{D}_\alpha[\hat{\rho}]\equiv\sum_{\alpha=\uparrow,\downarrow}\gamma_\alpha(R)\Bigl(\hat{L}_\alpha \hat{\rho}\hat{L}^+_\alpha-\frac{1}{2}\{\hat{L}^+_\alpha \hat{L}_\alpha,\hat{\rho}\}\Bigr).
\end{equation}
$\gamma_\alpha(R)(>0)$ is the coefficient of dissipators, and 
$\hat{L}^+_\alpha$ and  $\hat{L}_\alpha$ defined by
\begin{equation} \label{jump}
\hat{L}^+_{\uparrow}=\hat{L}_{\downarrow}=\hat{a}, \;\;\;\;  \hat{L}_{\uparrow}=\hat{L}^+_{\downarrow}=\hat{a}^+ 
\end{equation}
are the transition operators connecting between the adjacent quantum levels in Eqs.(\ref{eigenenerg}) and (\ref{eigen state}).  In Eq.(\ref{updown dissip}) the term with coefficient $\gamma_{\downarrow}(R) (\gamma_{\uparrow}(R)) $ describes  the damping (pumping) process.

We here assume the detailed balance condition to be explained below.  
When $R$ is constant, the solution  $\hat{\rho}(t, R)$ of GKSL equation in Eq.(\ref{GKSL-orig}) at $t\to+\infty$ tends to the fixed point of the dissipative dynamical system, which corresponds to
the Gibbs canonical distribution or Gibbs state, 
\begin{equation}\label{orig-Gibbs}
    \hat{\rho}(t=\infty, R)\equiv  \hat{\rho}_{eq}(R)=Z^{-1}\exp{(-\beta\hat{H}_0(R))}
\end{equation}
with the partition function $Z\equiv {\rm Tr} e^{-\beta\hat{H}_0(R)}$.  It should be noted that $\hat{\rho}_{eq}(R)$ has no explicit time dependence. With use of the eigenstate bases in Eq.(\ref{eigen state}),  the Gibbs canonical distribution is expressed as
\begin{equation}\label{Gibbs-expl}
   \hat{\rho}_{eq}(R)=\sum_{n=0}^{\infty} |n\rangle f_n(R) \langle n|    
\end{equation}
with $f_n(R)$ defined by
\begin{eqnarray}\label{f-distr}
f_n(R)&\equiv&\langle n| \hat{\rho}_{eq}(R)|n \rangle\nonumber\\
&=&e^{-\beta E_n(R)} / \sum_{k=0}^{\infty} e^{-\beta E_k(R)} \nonumber\\
&=&e^{-\beta n \hbar \omega(R)} - e^{-\beta (n+1) \hbar \omega(R)}.
\end{eqnarray}
$ \hat{\rho}_{eq}(R)$ satisfies Eq.(\ref{GKSL-orig}), because it is commutable with $H_0$, and guarantees $\hat{D}_0[\hat{\rho}_{eq} ]=0$ under the detailed balance condition (DBC):
\begin{equation}\label{KMS-0}
\gamma_{\uparrow}(R)-\gamma_{\downarrow}(R) e^{-\beta\hbar \omega(R)}=0.
\end{equation}
Historically, DBC is traced back to Kubo-Martin-Schwinger (KMS) condition \cite{Kubo,Martin,Yablon} for the inverse-Fourier transforms of any two-point correlation function in the infinite system at the Gibbs state.

After the system has reached the Gibbs state, let the confining length $R$  or frequency $\omega$ be time-dependent. Then the Gibbs distribution   
$\hat{\rho}_{eq}$ in Eq.(\ref{orig-Gibbs})  or Eq.(\ref{Gibbs-expl}) with Eq.(\ref{f-distr}) cannot satisfy the GKSL equation as it stands.  
But the idea of FF provides a protocol which keeps  $\hat{\rho}_{eq}(R(t))$ as the solution of the innovated GKSL equation under $R=R(t)$.

The first half of FF scheme is the regularization procedure and the second one is the time rescaling to construct the FF variant of the GKSL equation.

To begin with, let $R(t)$ vary slowly and linearly in time as
\begin{equation}\label{adiab para}
    R(t)\equiv R_0+\varepsilon t
\end{equation}
with the constant $|\varepsilon|\ll1$, which means that a very long time of $O\Bigl(\frac{1}{|\varepsilon|}\Bigr)$ is required to see a recognizable change of $R(t)$ in Eq.(\ref{adiab para}).
We then introduce the regularized distribution  $\hat{\rho}^{reg}$ which has the same form as in Eq.(\ref{orig-Gibbs}):
\begin{equation}\label{reg-Gibbs}
    \hat{\rho}^{reg}=\hat{\rho}_{eq}(R(t))=Z^{-1}(R(t))e^{-\beta\hat{H}_0(R(t))}.
\end{equation}
At the same time, we prepare Hilbert space spanned by the basis functions $\{\phi_n(x,R(t)) = \langle x |n(R)\rangle\}$ which are the instantaneous eigenstates of $\hat{H}_0(R(t)) |n(R)\rangle=E_n(R(t))|n(R)\rangle$.

We assume $\hat{\rho}^{reg}$ to satisfy the regularized GKSL equation where the Hamiltonian has an additional term  as $\hat{H}_0+\varepsilon \hat{h}$  and coefficients of Lindblad dissipators include an extra term $\varepsilon g_{\alpha}(R)$ as 
\begin{eqnarray}\label{DBE-A}
 \gamma_{\uparrow}(R)+\varepsilon g_{\uparrow}(R), \nonumber\\
 \gamma_{\downarrow}(R)+\varepsilon g_{\downarrow}(R).
 \end{eqnarray}

The regularized GKSL equation is given by
\begin{equation}\label{r-GKSL}
    \frac{d\hat{\rho}^{reg}}{dt}=-\frac{i}{\hbar}\Bigl[\hat{H}_0+\varepsilon\hat{h},\hat{\rho}^{reg}\Bigr]+\hat{D}_0[\hat{\rho}^{reg}]
    +\varepsilon \hat{d}[\hat{\rho}^{reg}]
\end{equation}
with
\begin{eqnarray}\label{correct-dissip}
\hat{d}[\hat{\rho}^{reg}] &\equiv& \hat{d}_{\uparrow} [\hat{\rho}^{reg}] + \hat{d}_{\downarrow} [\hat{\rho}^{reg}] \nonumber\\
&\equiv&\sum_{\alpha=\uparrow,\downarrow}g_{\alpha}(R)\Bigl(\hat{L}_\alpha \hat{\rho}^{reg}\hat{L}^+_\alpha-\frac{1}{2}\{\hat{L}^+_\alpha \hat{L}_\alpha,\hat{\rho}^{reg}\}\Bigr).\nonumber\\
\end{eqnarray}
Thanks to the relation $[\hat{H}_0,\hat{a}]=-\hbar\omega\hat{a} $ and $ [\hat{H}_0,\hat{a}^+]=\hbar\omega\hat{a}^+$, we see $Z^{-1}\hat{a}e^{-\beta\hat{H}_0}=Z^{-1}e^{-\beta(\hat{H}_0+\hbar\omega)}\hat{a}$ and $Z^{-1}\hat{a}^+e^{-\beta\hat{H}_0}
=Z^{-1}e^{-\beta(\hat{H}_0-\hbar\omega)}\hat{a}^+$. Then Eq.(\ref{correct-dissip}) becomes
\begin{eqnarray}\label{correct-dissip-final}
\hat{d}[\hat{\rho}^{reg}(R)] &=&(g_{\uparrow}(R)-g_{\downarrow}(R)e^{-\beta\hbar\omega})\hat{\rho}^{reg}(R)\nonumber\\
&\times&\Bigl(e^{\beta\hbar\omega}\hat{a}^+\hat{a}-\hat{a}\hat{a}^+\Bigr) .
\end{eqnarray}

Noting that $\hat{\rho}^{reg}(R)$ depends on time ($t$) only through $R(t)$ due to the definition in Eq.(\ref{reg-Gibbs}) with Eq.(\ref{orig-Gibbs}), we shall expand both sides of Eq.(\ref{r-GKSL}) w.r.t. $\epsilon$, and obtain successively the equalities of $O(1)$ and $O(\epsilon)$.
Using $\frac{d \hat{\rho}^{reg}(R)}{dt}=\frac{\partial \hat{\rho}^{reg}(R)}{\partial t}+\epsilon\frac{\partial \hat{\rho}^{reg}(R)}{\partial R}$,
the equality of $O(1)$ in Eq.(\ref{r-GKSL}) is:
\begin{equation}\label{eqlty-0}
\frac{\partial \hat{\rho}^{reg}(R)}{\partial t}=-\frac{i}{\hbar}\Bigl[\hat{H}_0,\hat{\rho}^{reg}\Bigr]+\hat{D}_0[\hat{\rho}^{reg}]=0.
\end{equation}
Then the equality of $O(\varepsilon)$ is given by 
\begin{equation}\label{core0}
\frac{\partial\hat{\rho}^{reg}(R)}{\partial R}=-\frac{i}{\hbar}[\hat{h},\hat{\rho}^{reg}(R)]+\hat{d}[\hat{\rho}^{reg}(R)].
\end{equation}
It is obvious that $\hat{\rho}^{reg}(R)$ satisfies Eq.(\ref{eqlty-0}): Firstly $[\hat{H}_0,\hat{\rho}^{reg}]=0$, and
secondly, $\hat{D}_0[\hat{\rho}^{reg}]=\hat{D}_\downarrow[\hat{\rho}^{reg}]+\hat{D}_\uparrow[\hat{\rho}^{reg}]=0$ because of the DBC in Eq.(\ref{KMS-0}) under the prescription in Eq.(\ref{jump}). 
Eq. (\ref{core0}) is the core equation of the inverse problem, and reveals the protocol $\hat{h}$ and $\hat{d}$ which guarantees  
the quasi-adiabatic Gibbs distribution $\hat{\rho}^{reg}$ in Eq.(\ref{reg-Gibbs}) under $R=R(t)$ in Eq.(\ref{adiab para}) after preparing in advance the Gibbs state with use of $R=R_0$.

\section{Nonadiabatic additional Hamiltonian and dissipator}\label{NAHD}
To obtain $\hat{h}$ and $d$, let us have recourse to the matrix elements of Eq.(\ref{core0}) using the adiabatic eigenstates in Eq.(\ref{eigen state})
with $R=R(t)$ in Eq.(\ref{adiab para}).
$(l, n)$ element of the left-hand side of Eq.(\ref{core0}) becomes
\begin{eqnarray}   
\langle l| \frac{\partial\hat{\rho}^{reg}}{\partial R} |n\rangle =\frac{\partial}{\partial R}\langle l|\hat{\rho}^{reg} |n\rangle \nonumber\\
-\Bigl(\frac{\partial}{\partial R}\langle l|\Bigr)\hat{\rho}^{reg}|n\rangle-\langle l|\hat{\rho}^{reg}\Bigl(\frac{\partial}{\partial R}|n\rangle\Bigr) \nonumber\\
=\delta_{ln}\frac{\partial}{\partial R}f_n 
+\langle l|\Bigl(\frac{\partial}{\partial R}|n\rangle\Bigr) (f_n - f_l),
\end{eqnarray}
where we used Eq.(\ref{f-distr}) and $\Bigl(\frac{\partial}{\partial R}\langle l|\Bigr)|n\rangle+\langle l|\Bigl(\frac{\partial}{\partial R}|n\rangle\Bigr)
= \frac{\partial}{\partial R} \langle l|n\rangle=0$.
The 1st and 2nd matrix elements of the right-hand side of Eq.(\ref{core0}) are
\begin{eqnarray}
\langle l|[\hat{h},\hat{\rho}^{reg}]|n\rangle=\sum_{k=0}^{\infty}(h_{lk}\rho_{kn}-\rho_{lk}h_{kn}) =h_{ln}(f_n-f_l) \nonumber\\
\end{eqnarray}
and
\begin{equation}   
\langle l|d[\hat{\rho}^{reg}] |n\rangle=\delta_{ln}\langle n|d[\hat{\rho}^{reg}] |n\rangle,
\end{equation}
respectively.

Comparing off-diagonal ($l \ne n$) matrix elements of both sides of Eq.(\ref{core0}), 
we find one equality:
\begin{equation}\label{h-Hamilt}
\langle l|\Bigl(\frac{\partial}{\partial R}|n\rangle\Bigr) =-\frac{i}{\hbar} h_{ln}.
\end{equation}

Similarly, comparison of the diagonal($l = n$) elements gives another equality
\begin{equation}\label{d-dissip}
   \frac{\partial}{\partial R}f_n=\langle n|d[\hat{\rho}^{reg}] |n\rangle.
\end{equation}

Using Eq.(\ref{eigen state}), both sides of Eq.(\ref{h-Hamilt}) are calculable:
its left-hand side becomes 
$\langle l|\Bigl(\frac{\partial}{\partial R}|n\rangle\Bigr)=\int_{-\infty}^{\infty}dx\phi_l^*(x)\frac{\partial}{\partial R}\phi_n(x)=-\frac{1}{2}\sqrt{\frac{m\omega}{\hbar}}\Bigl(\sqrt{n(n-1)}\delta_{l,n-2}-\sqrt{(n+1)(n+2)}\delta_{l,n+2}\Bigr)$. The corresponding right hand side is given with choice of $\hat{h}=a(\hat{p}\hat{x}+\hat{x}\hat{p})$ as
$h_{ln}=\langle l|a(\hat{p}\hat{x}+\hat{x}\hat{p}) |n\rangle=-ia\hbar\Bigl(\sqrt{n(n-1)}\delta_{l,n-2}-\sqrt{(n+1)(n+2)}\delta_{l,n+2}\Bigr)$.
Applying the equality in Eq.(\ref{h-Hamilt}), we can determine
 $a=\frac{1}{2}\sqrt{\frac{m\omega}{\hbar}}$ and obtain
\begin{equation}\label{h-fix}
\hat{h}=\frac{1}{2}\sqrt{\frac{m\omega}{\hbar}}(\hat{p}\hat{x}+\hat{x}\hat{p}).
\end{equation}

On the other hand, both sides of Eq.(\ref{d-dissip}) are calculable. Its left-hand side is
given by $\frac{\partial}{\partial R}f_n=\beta\hbar\frac{\partial\omega}{\partial R}((n+1) -ne^{\beta\hbar\omega}) e^{-\beta(n+1)\hbar\omega}$ with use of the definition of $f_n$ in Eq.(\ref{f-distr}). The corresponding right-hand is 
$\langle n|d[\hat{\rho}^{reg}] |n\rangle
=\langle n| (g_{\uparrow}(R)-g_{\downarrow}(R)e^{-\beta\hbar\omega})\hat{\rho}^{reg}(R)
\Bigl(e^{\beta\hbar\omega}\hat{a}^+\hat{a}-\hat{a}\hat{a}^+\Bigr)  |n\rangle=(g_{\uparrow}(R)-g_{\downarrow}(R)e^{-\beta\hbar\omega})
f_n(-(n+1)+ne^{\beta\hbar\omega})$ .
Equating both sides, we find
\begin{eqnarray}\label{extra dissip0}
g_{\uparrow}(R)-g_{\downarrow}(R)e^{-\beta\hbar\omega}=-\frac{\partial \omega}{\partial R}\beta\hbar e^{-\beta (n+1)\hbar \omega}/f_n \nonumber\\
=2\beta\frac{\hbar^2}{m}R^{-3}/(e^{\beta\hbar\omega}-1),
\end{eqnarray}
which shows the nonadiabatic correction to the standard DBC.
With use of an arbitrary constant $c$, one may rewrite the above result explicitly as
\begin{equation}\label{explicit g}
 \begin{pmatrix} g_{\uparrow}(R)\\ g_{\downarrow}(R) \end{pmatrix} =2\beta\frac{\hbar^2}{m}R^{-3}
 \frac{e^{\beta\hbar\omega}}{(e^{\beta\hbar\omega}-1)^2} \begin{pmatrix} 1+c \\ 1+e^{\beta\hbar\omega} c\end{pmatrix}.
\end{equation}
In this way, we have solved the inverse problem to reveal the protocols $\hat{h}$ and $d$ which keep the Gibbs distribution under an adiabatic change of the parameter $R$. The next and final subject is the FF time rescaling to realize a rapid change of the confining length $R$.

\section{Fast forward-and-backward time rescaling and time-periodic Gibbs state}\label{FF rescaling}
The regularized GKSL equation in Eq.(\ref{r-GKSL}) and its solution $\hat{\rho}^{reg}$ in Eq.(\ref{reg-Gibbs}) work well in the case that the confining length 
$R(t)$ changes extremely slowly and is monotonously increasing, which is not convenient for experimentalists who want to see the arbitrary time evolution of the control parameter. This problem can be resolved by extending our idea of fast forward, namely by replacing the time variable $t$ of $R(t)$ appearing in Eqs.(\ref{adiab para}) - (\ref{correct-dissip-final})  by its fast forward-and-backward variant $\Lambda(t)$ generated respectively by 
a very large positive and negative time-scaling factor $\alpha(t)$ as
\begin{equation}
\Lambda(t)=\int_0^{t}\alpha(t') dt'.
\label{future}
\end{equation}
After the above time scaling, the confinement parameter $R$
varies in time rapidly as
\begin{equation}
R( \Lambda(t))=R_0+\epsilon \Lambda(t).
\end{equation}

Below we shall employ a trick of  the infinitely-large factor $\alpha(t)$ to compensate the infinitesimally-small adiabaticity $\epsilon$, and
construct the time-periodic function $R(\Lambda(t))$ with its period $T_{FF}$ which is an arbitrary-short laboratory time.
In the range $0\leq t \leq T_{FF} $, we define $\alpha(t)$ as
$\alpha(t)=1+ \bar{\alpha}f(s)$ with $s \equiv \frac{t}{T_{FF}}$, where  $\bar{\alpha} (\gg 1)$ satisfies the asymptotic limit
$ \lim_{\epsilon\to 0, \bar{\alpha} \to \infty} \epsilon \bar{\alpha} = \bar{v} (=O(1))$. 
The mean velocity $\bar{v}$ characterizes the speed of FF periodic driving.  $f(s)$ is firstly defined in the range $0 \le s \le 1$ and satisfies $\bar{f}= \int_0^1f(s')ds'=0$ so that $R(\Lambda(t))$ recovers the original value $R_0$. Secondly
 $f(s)$ is made periodic as $f(s+1)=f(s)$ with
$f(0)=f(1)=f(2)=\cdots=0$. The functional form of $f(s)$ is completely free except for this boundary condition.
Among a wide family of functions of $f(s)$, we choose the simplest function $\sin(2\pi s)$. Then
\begin{equation}
\alpha(t) =1+\bar{\alpha} \sin(\frac{2 \pi}{T_{FF}}t).
\label{alp-scal}
\end{equation}

In the asymptotic limit we can introduce
the the velocity function as
\begin{eqnarray}\label{vt}
v(t) = \lim_{\epsilon\to 0, \bar{\alpha} \to \infty} \epsilon \alpha(t) = \bar{v}\sin(\frac{2 \pi}{T_{FF}}t),
\end{eqnarray}
and then the confinement parameter can change radically and periodically as
\begin{eqnarray}\label{lambda2}
R(\Lambda(t))&=&R_0+\lim_{\epsilon\rightarrow 0, \bar{\alpha}\rightarrow \infty}\varepsilon\Lambda(t) 
=R_{0}+\int^{t}_{0}v(t')dt'\nonumber\\
&=& R_{0}+\frac{\bar{v}T_{FF}}{2\pi}\left[1-\cos\left(\frac{2\pi}{T_{FF}}t\right)\right].
\end{eqnarray}
For simplicity we employ
\begin{eqnarray}\label{mean veloc}
\frac{\bar{v}T_{FF}}{R_0}=1,
\end{eqnarray}
although this ratio is an arbitrary constant of $O(1)$. 

Let us now define the FF Gibbs distribution $\hat{\rho}_{FF}$ as 
\begin{eqnarray}\label{FF Gibbs}
\hat{\rho}_{FF} &\equiv &\hat{\rho}(t=\infty, R(\Lambda(t)))=\hat{\rho}^{reg}(R(\Lambda(t))), \nonumber\\
&=&Z^{-1}(R(\Lambda(t))) e^{-\beta H_0(R(\Lambda(t)))},
\end{eqnarray}
indicating that $\hat{\rho}_{FF}$ is time-periodic and depends on time $t$  implicitly through $R(\Lambda(t))$.
$\hat{\rho}_{FF}$ proves to satisfy the same GKSL equation as Eq.(\ref{r-GKSL}) with $\epsilon$ prior to $\hat{h}$ and $d$ being replaced by 
$\epsilon \alpha$ and $R(t)$ included in $\hat{H}_0, \hat{h}, D_0$, and $d $ by $R(\Lambda(t))$.
In fact, the time derivative of $\hat{\rho}_{FF}$ is a sum of its explicit and implicit derivatives:
\begin{eqnarray}\label{R-Lamda}
\frac{d \hat{\rho}_{FF}}{dt}=\frac{\partial \hat{\rho}_{FF}}{\partial t}+\frac{dR}{dt}\frac{\partial \hat{\rho}_{FF}}{\partial R}.
\end{eqnarray}
On the r.h.s. of  Eq.(\ref{R-Lamda}), the 1st term is equal to the r.h.s. of Eq.(\ref{eqlty-0}) with $\hat{\rho}^{reg}$ replaced by $\hat{\rho}_{FF}$,
and the 2nd term is given by the FF variant of Eq.(\ref{core0}) multiplied 
with $\epsilon\alpha(=\frac{dR}{dt}=\frac{dR}{d\Lambda}\frac{d\Lambda}{dt})$.

In the asymptotic limit under consideration,
we obtain GKSL equation working for the rapid-time scale region:
\begin{eqnarray}
    \frac{d\hat{\rho}_{FF}}{dt}&=&-\frac{i}{\hbar}\Bigl[\hat{H}_0+v(t)\hat{h},\hat{\rho}_{FF}\Bigr] \nonumber\\
    &+&\hat{D}_0[\hat{\rho}_{FF}]+v(t)\hat{d}[\hat{\rho}_{FF}].
\label{FF-GKSL}
\end{eqnarray}

From now on we take the following prescription:
\begin{eqnarray} \label{R-FF}
R(t) &\equiv &R(\Lambda(t)),\nonumber\\
 \dot{R} (\equiv  \frac{dR}{dt} )& \equiv &\frac{dR(\Lambda(t))}{dt}=v(t) .
\label{new-kap}
\end{eqnarray}
Then we see $R(t)$ to be a periodic function of $t$ with the period $T_{FF}$, satisfying 
\begin{eqnarray}\label{bound}
R(nT_{FF})=R_0, \qquad \dot{R}(nT_{FF})=0
\end{eqnarray}
with $n=0,1,2,\cdots$.

As noted above Eq.(\ref{alp-scal}), the idea of FF accepts a wide family of functions which satisfy the above boundary condition.
Besides the trigonometric function in Eq.(\ref{lambda2}), for example, one may employ a polynomial function. If we choose $f(s)=2s(s-1)(2s-1)$ in the range
$0 \le s \le 1$ and let it periodic by the definition $f(s +n)=f(s)$ with $n=1,2,3,\cdots$. Then $\dot{R}(t)=v(t)=2\bar{v}\frac{t}{T_{FF}}(\frac{t}{T_{FF}}-1)
(2\frac{t}{T_{FF}}-1)$
and
\begin{eqnarray}\label{lambda3}
R(t)=R_0&+&\bar{v}T_{FF}\left[ (\frac{t}{T_{FF}})^2(\frac{t}{T_{FF}}-1)^2\right] \nonumber\\
&& (0\le t<T_{FF}),\nonumber\\
R(t+nT_{FF})&=&R(t)    \qquad(n=1,2,3, \cdots).
\end{eqnarray}

To summarize, the expression of the Gibbs state in Eq.(\ref{FF Gibbs}) is given by
\begin{eqnarray}\label{Gibbs rev}
\hat{\rho}_{FF}(R)=\sum_{n=0}^{\infty} |n(R)\rangle f_n(R)\langle n(R)| 
\end{eqnarray}
and the FF variant of GKSL equation takes the following form:
\begin{eqnarray}\label{FF-GKSL}
    \frac{d\hat{\rho}_{FF}}{dt}&=&-\frac{i}{\hbar}\Bigl[\hat{H}_{FF},\hat{\rho}_{FF}\Bigr] +\hat{D}_{FF}[\hat{\rho}_{FF}]
\end{eqnarray}
with
\begin{eqnarray}\label{FF-Haml}
\hat{H}_{FF}&\equiv&\hat{H}_0+\dot{R}\hat{h}=\frac{\hat{p}^2}{2m}+\frac{\hbar^2}{2m R^4}\hat{x}^2\nonumber\\
&+&\frac{\dot{R}}{2R}(\hat{p}\hat{x}+\hat{x}\hat{p}).
\end{eqnarray}
The FF dissipator in Eq.(\ref{FF-GKSL}) consists of the standard $\hat{D}_0[\hat{\rho}_{FF}]$ and additional $\dot{R}\hat{d}[\hat{\rho}_{FF}]$ parts as
\begin{eqnarray}\label{FF-Dissip}
&&\hat{D}_{FF}[\hat{\rho}_{FF}]\equiv \hat{D}_0[\hat{\rho}_{FF}]+\dot{R}\hat{d}[\hat{\rho}_{FF}]\nonumber\\
&=&\sum_{\alpha=\uparrow,\downarrow}\gamma^{FF}_\alpha(R, \dot{R})\Bigl(\hat{L}_\alpha \hat{\rho}_{FF}\hat{L}^+_\alpha-\frac{1}{2}\{\hat{L}^+_\alpha \hat{L}_\alpha,\hat{\rho}_{FF}\}\Bigr),\nonumber\\
\end{eqnarray}
where
\begin{eqnarray}\label{FF-dissi coeffs}
\gamma^{FF}_\alpha(R, \dot{R})\equiv \gamma_\alpha(R)+\dot{R}g_\alpha(R). 
\end{eqnarray}

Equation (\ref{R-Lamda}) decomposed the time derivative of the density matrix into “explicit” and “implicit” parts after time rescaling. The explicit derivative of the density matrix  can be vanishing because it is a fixed-point solution of Eq.(\ref{eqlty-0}).  The implicit derivative with use of the chain rule has given $\dot{R}\hat{h}$ and  
$\dot{R}\hat{d}$ in Eqs.(\ref{FF-Haml}) and (\ref{FF-Dissip}), respectively.

\begin{figure}[!h]
\begin{center}
\includegraphics[width=3.0in]{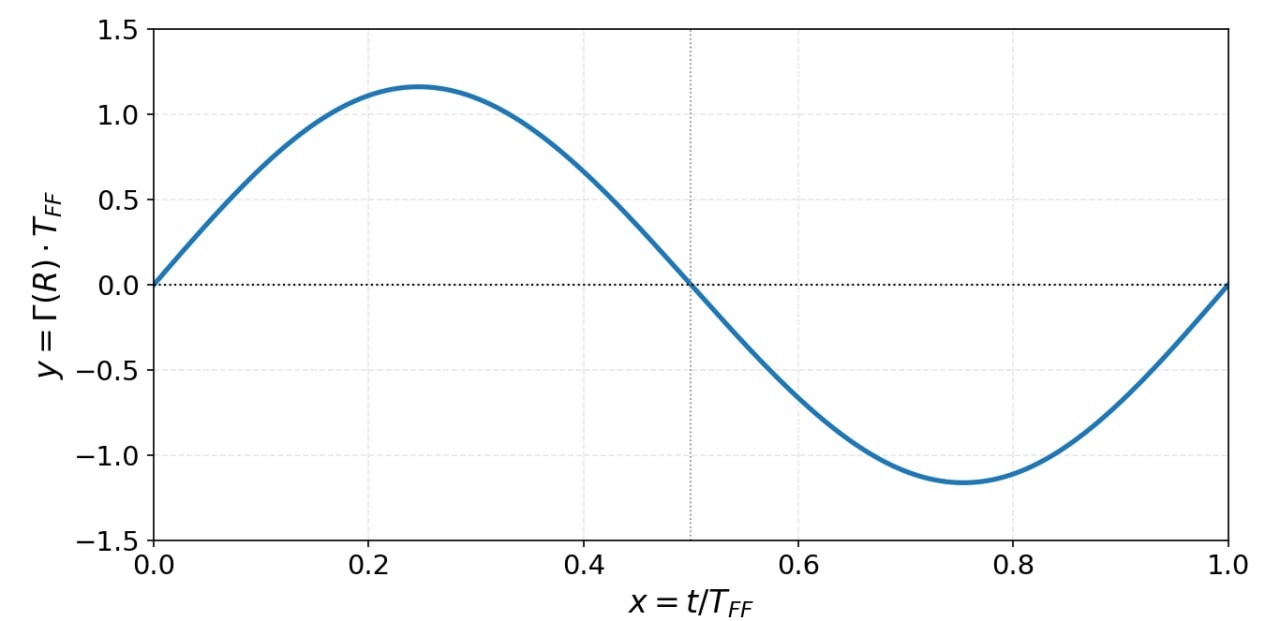}
\includegraphics[width=3.0in]{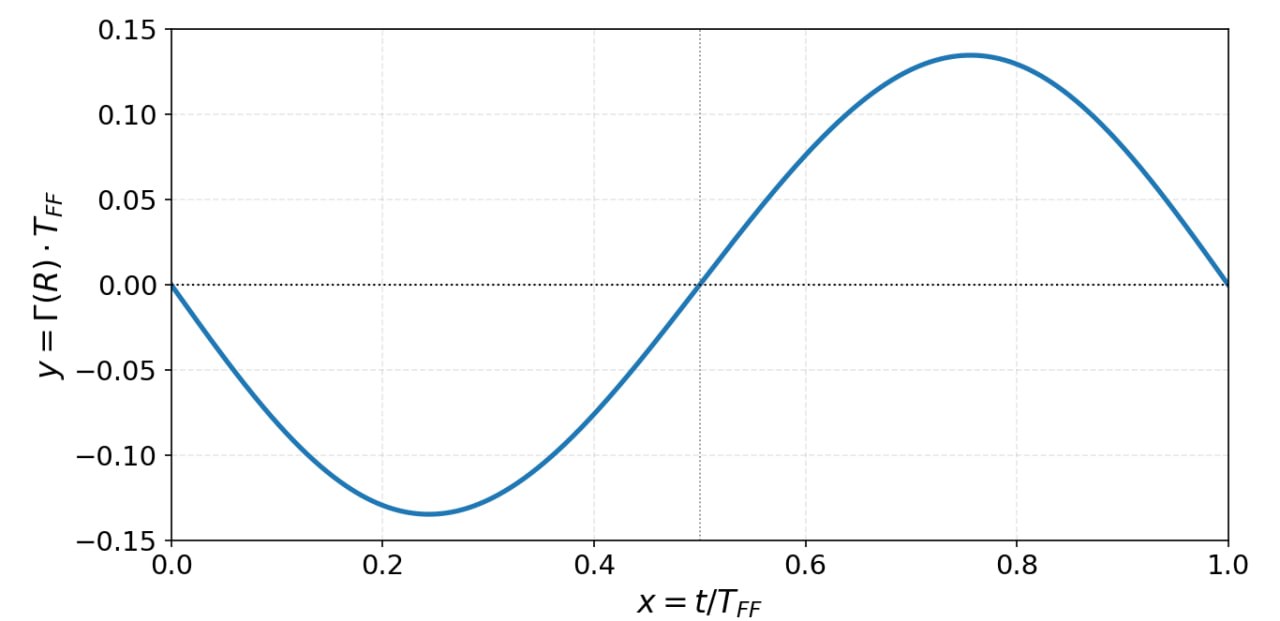}
\caption{Time dependence of FF variant of DBC, $\Gamma(R, \dot{R})$, in the one period $0\le t\le T_{FF}$. 
$\dot{R}=v(t)$ in Eq.(\ref{vt}), $R(\Lambda(t))$ in Eq. (\ref{lambda2}) and $\bar{v}T_{FF}/R_0=1$ in Eq.(\ref{mean veloc}) are employed.
Upper panel for harmonic oscillator:  $\Gamma(R, \dot{R})$ in Eq.(\ref{DBE-Coeff}) with $\frac{\beta\hbar^2}{mR_0^2}=1$; 
Lower panel for spin: $\Gamma(R, \dot{R})$ in Eq.(\ref{FF DBC}) with $\beta R_0=1$ and $\frac{B_0}{ R_0}=1$.
Zero lines with $y=0$ correspond to the standard DBCs (i.e., KMS conditions) in Eqs. (\ref{KMS-0}) and (\ref{orig DBC}) 
in the case of constant $R$.} \label{FF time scal}
\end{center}
\end{figure}

Because of the extra coefficient $\dot{R}g_\alpha(R)$, the detailed balance condition (DBC) in Eq.(\ref{KMS-0}) is now replaced by its FF variant as
\begin{eqnarray}\label{DBE-Coeff}
\gamma^{FF}_{\uparrow}(R, \dot{R})&-&\gamma^{FF}_{\downarrow}(R, \dot{R})e^{-\beta\hbar\omega(R)}\nonumber\\
&=&\dot{R}(g_{\uparrow}(R)-g_{\downarrow}(R)e^{-\beta\hbar\omega(R)})\nonumber\\
&=&2\beta\frac{\hbar^2}{m}\dot{R}R^{-3}/(e^{\beta\hbar\omega}-1)\equiv \Gamma(R,\dot{R}) ,\nonumber\\
\end{eqnarray}
where equalities in Eqs. (\ref{KMS-0}) and (\ref{extra dissip0}) are used.


Figure \ref{FF time scal} (upper panel) shows the behavior of $\Gamma(R)$ as a function of $t$ in the one-period range $0\leq t \le T_{FF}$.
Thanks to the FF correction $\dot{R}\hat{h}$ to the original Hamiltonian $\hat{H}_0$ and the extra dissipator of Lindblad type with coefficient in Eq.(\ref{DBE-Coeff}), the FF Gibbs state in Eq.(\ref{FF Gibbs}) keeps to satisfy the GKSL equation in (\ref{FF-GKSL}). 
Regarding  $T_{FF}$ as a period of periodic driving like $R(t+T_{FF})=R(t)$ with $t=0$ chosen as the starting time of protocol, we can see creation of the time-periodic Gibbs state despite a rapid change of the frequency of HO under the FF protocol.

There exists a related work on STA with designing the environments \cite{Campo1} 
in the context of quantum master equations. The relationship of dissipators between STA and FF approaches is briefly explained in Appendix \ref{apdB}.

\section{FF scheme applied to a driven dissipative quantum spin system}\label{AOQD2}
To see a wide applicability of the FF scheme explained so far, we choose another prototype, i.e., a driven quantum spin system and show the FF variant of Hamiltonian and detailed balance condition
which supports the non-equilibrium Gibbs state under the time-periodic magnetic field. 

Let's introduce a dissipative quantum spin system 
under the applied magnetic field $\mathbf{B} \equiv (B_x, B_y, B_z ) = (R, 0, B_0)$, where
$B_0$ is fixed and $R$ will be taken as an external control parameter. The corresponding Hamiltonian $\hat{H}_0(R)$ is expressed as
\begin{equation}\label{hamil-0}
\hat{H}_0(R) = \frac{1}{2} \mathbf{\hat{\sigma}} \cdot \mathbf{B} = \frac{1}{2} \begin{pmatrix} B_0& R\\ R & -B_0 \end{pmatrix} .
\end{equation}
This model comes from the scheme of magnetic resonance of a spin under the applied vertical field $B_z $ and perpendicular rotating field $B_x = R \cos(\omega t)$ and $B_y = R \sin(\omega t) $. Moving to the frame rotating around the $z$-axis, the system's Hamiltonian is equivalent to Eq. (\ref{hamil-0}) by introducing $B_0 = B_z - \omega/\gamma$ with the gyromagnetic ratio \( \gamma \).

The spin dynamics in the presence of the dissipative environment, i.e., a heat reservoir, is described by the Lindblad equation for 
the system’s density operator $\hat{\rho}$ as in Eq.(\ref{GKSL-orig})
where the dissipator $ \hat{D}_0[\hat{\rho}]$ is now written with use of jump operators defined between 
the ground and excited states.

In the case of constant $R(\ne 0)$ it is convenient to change the quantization axis of Pauli operators from $ \mathbf{e}_z$  to $\mathbf{e}_n$, which is parallel to the ground-state of eigenstates of $\hat{H}_0(R)$ in Eq.(\ref{hamil-0}).
$\mathbf{e}_n$ is described as $\mathbf{e}_n \equiv (n_x, n_y, n_z)=(\sin\theta \cos\phi, \sin\theta \sin\phi, \cos\theta )$. 
$\hat{\sigma}\cdot \mathbf{e}_n$ has eigenvectors
\begin{eqnarray}\label{tilde vect}
|\tilde{\uparrow} \rangle &=&-e^{-i\phi} \sin\left(\frac{\theta}{2}\right)|\downarrow \rangle+ \cos\left(\frac{\theta}{2}\right)|\uparrow \rangle, \nonumber\\
|\tilde{\downarrow} \rangle&=& e^{-i\phi}\cos\left(\frac{\theta}{2}\right)|\downarrow \rangle + \sin\left(\frac{\theta}{2}\right)|\uparrow \rangle, 
\end{eqnarray}
corresponding to eigenvalues \(-1\) and \(+1\), respectively,
where $\phi = 0$, and
\begin{equation}\label{relation}
\sin\theta = \frac{R}{\sqrt{R^2 + B_0^2}}, \quad
\cos\theta = \frac{B_0}{\sqrt{R^2 + B_0^2}}.
\end{equation}
The Pauli operators quantized about \( \mathbf{e}_n \) is denoted by \( \hat{\tau}_i \, (i = x, y, z) \) with
\begin{equation}\label{tau}
\hat{\tau}_x =
\begin{pmatrix}
0 & 1 \\
1 & 0
\end{pmatrix},
\quad
\hat{\tau}_y =
\begin{pmatrix}
0 & -i \\
i & 0
\end{pmatrix}
 ,
\quad
\hat{\tau}_z =
\begin{pmatrix}
1 & 0 \\
0 & -1
\end{pmatrix}
.
\end{equation}
In constructing the matrices in Eq.(\ref{tau}), we have employed the new bases $|\tilde{\downarrow}\rangle$ and $|\tilde{\uparrow}\rangle$ 
which are $R-$ and 
$B_0-$dependent through $\theta$ with $\phi=0$, as seen in Eqs.(\ref{tilde vect}) and (\ref{relation}).
The representation transformation of Pauli operators from $\{\hat{\sigma}_i\}$ to $\{\hat{\tau}_i\}$  is:
$\hat{\tau}_x\equiv |\tilde{\uparrow}\rangle \langle \tilde{\downarrow}|+ |\tilde{\downarrow}\rangle \langle \tilde{\uparrow}|
=\cos\theta\hat{\sigma}_x -\sin\theta \hat{\sigma}_z$, $\hat{\tau}_y \equiv i|\tilde{\uparrow}\rangle \langle \tilde{\downarrow}|- i|\tilde{\downarrow}\rangle \langle \tilde{\uparrow}|= \hat{\sigma}_y$ and
$\hat{\tau}_z\equiv |\tilde{\downarrow}\rangle \langle \tilde{\downarrow}|- |\tilde{\uparrow}\rangle \langle \tilde{\uparrow}|=\sin\theta\hat{\sigma}_x +\cos\theta \hat{\sigma}_z$.
Noting 
$\mathbf{B}= \sqrt{R^2 + B_0^2}(\frac{R}{\sqrt{R^2 + B_0^2}}, 0, \frac{B_0}{\sqrt{R^2 + B_0^2}})=\sqrt{R^2 + B_0^2} (\sin\theta, 0, \cos\theta)$,
Hamiltonian $\hat{H}_0(R)$ in Eq. (\ref{hamil-0}) is rewritten as
\begin{eqnarray}\label{sig tau}
\hat{H}_0(R) &=& \frac{1}{2} \sqrt{R^2 + B_0^2} \hat{\sigma} \cdot \bold{e}_n \nonumber\\
&=& \frac{1}{2} \sqrt{R^2 +B_0^2} \hat{\tau}_z
\end{eqnarray}
where the Hilbert space is now spanned with $|\tilde{\uparrow}\rangle$ and $|\tilde{\downarrow}\rangle$ in Eq. (\ref{tilde vect}).

On the other hand, the dissipator $\hat{D}_0[\rho]$ is expressed in terms of $\hat{\tau}$ operators as
\begin{eqnarray}\label{dissip 0}
\hat{D}_0[\hat{\rho}] &=& \gamma_\uparrow(R) \left( \hat{\tau}_+ \hat{\rho} \hat{\tau}_+^\dagger - \frac{1}{2} \{ \hat{\tau}_+^\dagger \hat{\tau}_+, \hat{\rho} \} \right)\nonumber\\
&+& \gamma_\downarrow(R) \left( \hat{\tau}_- \hat{\rho} \hat{\tau}_-^\dagger - \frac{1}{2} \{ \hat{\tau}_-^\dagger \hat{\tau}_-,\hat{ \rho} \} \right),
\end{eqnarray}
where \(\hat{\tau}_\pm = \frac{1}{2} (\hat{\tau}_x \pm i \hat{\tau}_y)\) are the Lindblad's jump operators connecting between $|\tilde{\uparrow}\rangle$ and 
$|\tilde{\downarrow}\rangle$. In Eq. (\ref{dissip 0}), $\gamma_\uparrow(R) (\gamma_\downarrow(R)) $ means the strength of pumping (damping) process.

Gibbs state $\hat{\rho}_{eq}(R)\equiv \frac{1}{Z(R)} e^{-\beta \hat{H}_0(R)}$ is expressed 
with use of eigenvalues $(\pm \frac{1}{2} \sqrt{R^2 + B_0^2})$ and eigenstates $(|\tilde{\uparrow}\rangle, 
|\tilde{\downarrow}\rangle)$ of $H_0(R)$ as 
\begin{eqnarray}\label{gibbs}
\hat{\rho}_{eq}(R) = \frac{1}{Z(R)} e^{-\beta \frac{\sqrt{R^2 + B_0^2}}{2}}|\tilde{\downarrow}\rangle \langle \tilde{\downarrow}|
+ \frac{1}{Z(R)} e^{\beta \frac{\sqrt{R^2 + B_0^2}}{2}}|\tilde{\uparrow}\rangle \langle \tilde{\uparrow}| \nonumber\\
\end{eqnarray}
with $Z(R)$ and \(\beta(=\frac{1}{k_BT})\) being the partition function and the inverse temperature, respectively.

When $R$ is constant, the solution of Lindblad equation becomes $\hat{\rho}_{eq}$ at $t \to \infty$, because $\hat{\rho}_{eq}$ commutes with $\hat{H}_0$ and guarantees $\hat{D}_0[\hat{\rho}_{eq}] = 0$ under the standard detailed balance condition (DBC) defined by
\begin{equation}\label{orig DBC}
\gamma_\uparrow(R) - \gamma_\downarrow(R) e^{-\beta \sqrt{R^2 + B_0^2}}=0.
\end{equation}

If $R$ will then become time-dependent,  $\hat{\rho}_{eq}$ cannot satisfy the Lindblad equation under the DBC in Eq.(\ref{orig DBC}). But, the idea of FF will provide a protocol which keeps $\hat{\rho}_{eq}(R(t))$ to satisfy the innovated Lindblad equation.

Assume $R(t)$ to vary adiabatically, i.e., slowly in time as in Eq.(\ref{adiab para})
with $|\epsilon| \ll 1$, and define the regularized distribution $\hat{\rho}^{\text{reg}}(R(t)) \equiv \hat{\rho}_{eq} (R(t))$.
By adding the corrections $\epsilon\hat{h}$ and $\epsilon\hat{d}$ respectively to Hamiltonian $\hat{H}_0$  and to the dissipator $\hat{D}_0$, 
$\hat{\rho}^{\text{reg}}(R(t))$ can keep the solution of the innovated Lindblad equation.
$\hat{h}$ and $\hat{d}$ can be obtained by solving the core equation as in Eq.(\ref{core0}), where
\begin{eqnarray}\label{deriv rho}
\frac{\partial \hat{\rho}^{\text{reg}}}{\partial R} &=&- \beta \frac{R}{\sqrt{R^2 + B_0^2}} \frac{e^{\beta\sqrt{R^2+B_0^2}}}{(e^{\beta\sqrt{R^2 + B_0^2}} + 1)^2}\cdot\hat{\tau}_z\nonumber\\
&-&\frac{1}{2} \frac{B_0}{ R^2+ B_0^2} \tanh \left( \beta \frac{\sqrt{R^2 + B_0^2}}{2} \right) \cdot \hat{\tau}_x. 
\end{eqnarray}
As in Section \ref{NAHD} for the harmonic oscillator, the off-diagonal and diagonal elements of Eq.(\ref{deriv rho})  determine $\hat{h}$ and $\hat{d}$, respectively.

Then, by replacing the time variable ($t$) by its advanced or future variant $\Lambda(t) \equiv \int_0^t \alpha(t') dt'$ with use of  a time-periodic large scaling factor 
$\alpha(t)$ competitive to the adiabaticity, the external parameter $R$ becomes $R(\Lambda(t)) = R_0 + \epsilon \Lambda(t)$, and we can define the FF variant of Gibbs state $\hat{\rho}_{\text{FF}}\equiv\hat{\rho}^{reg}(R(\Lambda(t)))$.

Taking the asymptotic limit $\lim_{\epsilon \to 0,\bar{\alpha} \to \infty} \epsilon \bar{\alpha}= \bar{v}$, we obtain a Lindblad-type equation with the innovated DBC working for the rapid-time scale region.
As in the case of the harmonic oscillator, the periodic driving $R(t)\equiv R(\Lambda (t))$ takes the same form as in Eqs. (\ref{lambda2}) or (\ref{lambda3}) with the period $T_{FF}$.

The FF of time-periodic Gibbs state, $\hat{\rho}_{\text{FF}}(R)$, is given by
\begin{eqnarray}\label{rho explicit}
 \hat{\rho}_{\text{FF}}(R)=\frac{1}{2}\big(\hat{I}_2  - \tanh\left(\frac{\beta}{2}\sqrt{R^2(t)+B_0^2}\right)\hat{\tau}_z\big)
\end{eqnarray}
and the FF variant of the Lindblad equation takes the form as
$\frac{d\hat{\rho}_{\text{FF}}}{dt} = -\frac{i}{\hbar} \left[ \hat{H}_{\text{FF}},\hat{\rho}_{\text{FF}} \right] + \hat{D}_{\text{FF}}[\hat{\rho}_{\text{FF}}]$
with
\begin{eqnarray}\label{FF Hamil2}
\hat{H}_{\text{FF}}& \equiv&\hat{H}_0+\dot{R} \hat{h} \nonumber\\
&=& \frac{1}{2}\sqrt{R^2(t)+B_0^2}\hat{\tau}_z+\frac{\hbar}{2}\frac{\dot{R}B_0}{R^2(t)+B_0^2}\hat{\tau}_y.
\end{eqnarray}
The FF dissipator consists of the standard and additional parts as
$\hat{D}_{FF}[\hat{\rho}_{FF}]\equiv \hat{D}_0[\hat{\rho}_{FF}]+\dot{R}\hat{d}[\hat{\rho}_{FF}]
=\sum_{\alpha=\uparrow,\downarrow}\gamma^{FF}_\alpha(R)\Bigl(\hat{L}_\alpha \hat{\rho}_{FF}\hat{L}^+_\alpha-\frac{1}{2}\{\hat{L}^+_\alpha \hat{L}_\alpha,\hat{\rho}_{FF}\}\Bigr)$
with $\hat{L}_{\uparrow}\equiv\hat{\tau}_+$, $\hat{L}_{\downarrow}\equiv\hat{\tau}_-$ and $\gamma_\alpha^{\text{FF}}(R, \dot{R}) = \gamma_\alpha(R) + \dot{R}g_\alpha(R)$.
Because of the extra coefficient $ \dot{R}g_\alpha(R)$, the detailed balance condition (DBC) is now generalized as
\begin{eqnarray}\label{FF DBC}
&&\gamma_{\uparrow}^{\text{FF}}(R, \dot{R}) -\gamma_{\downarrow}^{\text{FF}}(R, \dot{R}) e^{-\beta\sqrt{R^2(t)+B_0^2}}\nonumber\\
&=&-\beta\dot{R}R \frac{1}{\sqrt{R^2(t)+B_0^2}\Bigl(e^{\beta\sqrt{R^2(t)+B_0^2}}+1\Bigr)} \equiv \Gamma(R, \dot{R}).\nonumber\\
\end{eqnarray}
In this way, the FF scheme has resulted in the same issue as in the case of HO.

Figure \ref{FF time scal} (lower panel) shows the behavior of $\Gamma(R, \dot{R})$ as a function of $t$ in the range $0 \le t \le T_{\text{FF}}$ in the case $R\equiv R(\Lambda(t))$ in Eq.(\ref{lambda2}). In this figure, the reservoir temperature $\frac{1}{\beta}$ is chosen to be equal to the energy gap $B_0$. Thanks to the FF correction $\dot{R}\hat{h}$ to the original Hamiltonian $\hat{H}_0$ and the extra dissipation coefficients satisfying the FF variant of DBC in Eq.(\ref{FF DBC}), the periodic  Gibbs state in Eq.(\ref{rho explicit}) becomes a limit-cycle solution of the innovated GKSL equation above Eq. (\ref{FF Hamil2}). 




\section{Energy conservation, reversibility and constraint on the speed of FF protocol}\label{implement}
Recently, in the research area of open quantum systems,  there appeared interesting works on the quantum speed limit addressing a trade-off between the operation time and energy fluctuations \cite{funo}
and the trade-off relation between heat current and entropy production rate \cite{funo2}. In those works, however, coefficients of dissipators are assumed to satisfy the standard detailed balance or the KMS condition \cite{Kubo,Martin,Yablon} which guarantees the thermalization, i.e., realization of the Gibbs state. 
The present work is complementary to them. Starting from the Gibbs state thus prepared,
we modified dissipators together with Hamiltonian to keep the non-equilibrium Gibbs state under the FF periodic protocol.

Below, concentrating on the harmonic oscillator under the FF protocol, we shall see the FF variant of DBC in Eq.(\ref{DBE-Coeff}) to guarantee the energy conservation and reversibility, and also show a constraint on the speed of driving.
Let's investigate the time derivative of the work and heat in the FF evolution of Gibbs state assuming the isothermal process.
At any instant of time, the system's internal energy is defined by
\begin{eqnarray}\label{energ}
E={\rm Tr}(\hat{H}_{FF}\hat{\rho}_{FF}).
\end{eqnarray}
The incoming work current (: work done from the outside per time) and heat current (: heat flow from the reservoir per time) are given respectively by
\begin{eqnarray}\label{t-der work}
\dot{W}={\rm Tr}(\dot{\hat{H}}_{FF}\hat{\rho}_{FF}),
\end{eqnarray}
and
\begin{eqnarray}\label{t-der heat}
\dot{Q}={\rm Tr}(\hat{H}_{FF}\dot{\hat{\rho}}_{FF}).
\end{eqnarray}
The energy conservation (1st law of thermodynamics) during the FF protocol
\begin{eqnarray}\label{1st law}
\dot{W}+\dot{Q}=\dot{E}.
\end{eqnarray}
is available directly from the product rule, by taking the time derivative of Eq.(\ref{energ}).

Below, however, applying the FF variant of GKSL equation in 
Eq.(\ref{FF-GKSL}), we explicitly evaluate $\dot{W}$ and $\dot{Q}$.
From the definition of Eq.(\ref{energ}), we see:
\begin{eqnarray}\label{energ-cal}
E&=&\sum_{n=0}^{\infty} \langle n(R)|\hat{H}_{FF}|n(R)\rangle \langle n(R)|\hat{\rho}_{FF}|n(R)\rangle \nonumber\\
&=&\frac{1}{2}(\frac{\hbar^2}{mR^2})\coth(\beta\hbar\omega/2).
\end{eqnarray}
Here we used $\hbar \omega=\frac{\hbar^2}{mR^2}$ and the vanishing of the diagonal expectation value of the control term,
\begin{eqnarray}\label{nul-equality}
\langle n(R)|\dot{R}\hat{h}|n(R)\rangle=\frac{\dot{R}}{2R}\cdot\langle n(R)|(\hat{p}\hat{x}+\hat{x}\hat{p})|n(R)\rangle=0, \nonumber\\
\end{eqnarray}
which will also be exploited hereafter.
In the above derivation, the dissipator played no role.

On the other hand, noting that the matrix of $\hat{\rho}_{FF}$ in Eq.(\ref{Gibbs rev}) is diagonal, we find:
\begin{eqnarray}\label{t-der work}
\dot{W}&=&\sum_{n=0}^{\infty}\langle n(R)|\hat{\rho}_{FF}|n(R)\rangle \frac{\partial}{\partial t}\langle n(R)|\hat{H}_{FF}|n(R)\rangle \nonumber\\
&=&\frac{1}{2}\frac{\partial}{\partial t}\Bigl(\frac{\hbar^2}{mR^2}\Bigr)\coth(\beta\hbar\omega/2).
\end{eqnarray}
Similarly, $\dot{Q}$ is evaluated with use of the FF variant of DBC, $\Gamma(R)$ in Eq.(\ref{DBE-Coeff}), as
\begin{eqnarray}\label{t-der heat}
\dot{Q}&=&\sum_{n=0}^{\infty}\langle n(R)|\dot{\hat{\rho}}_{FF}|n(R)\rangle \langle n(R)|\hat{H}_{FF}|n(R)\rangle\nonumber\\
&=&\sum_{n=0}^{\infty} \Gamma(R) \langle n(R)|\hat{\rho}_{FF}(e^{\beta\hbar\omega}\hat{a}^+\hat{a}-\hat{a}\hat{a}^+) |n(R)\rangle\nonumber\\
&\times& \langle n(R)|\hat{H}_{FF}|n(R)\rangle \nonumber\\
&=&\frac{1}{2}\Bigl(\frac{\hbar^2}{mR^2}\Bigr)\frac{\partial}{\partial t}\coth(\beta\hbar\omega/2).
\end{eqnarray}
From Eqs.(\ref{t-der work}) and (\ref{t-der heat}), we understand: the work done on the system per unit time is positive ($\dot{W}>0$) for compression ($\dot{R}<0$), while the heat absorption per unit time is positive ($\dot{Q}>0$) for expansion($\dot{R}>0$).
The sum of  Eq.(\ref{t-der work}) and  Eq.(\ref{t-der heat}) gives the time derivative of energy $E$ in Eq.(\ref{energ-cal}), and we see that the FF variant of the detailed balance condition in Eq.(\ref{DBE-Coeff}) guarantees Eq.(\ref{1st law}), i.e., the 1st law of thermodynamics or energy conservation.

We shall then evaluate the entropy production rate ($\dot{S}$) of the system.  
Assuming the isothermal process, 
the system's entropy is given by $S=-{\rm Tr}(\hat{\rho}_{FF}\ln\hat{\rho}_{FF})$. 
Using ${\rm Tr}(\frac{d}{dt}\hat{\rho}_{FF})=0$, we see:
\begin{eqnarray}\label{t-der S entrop}
\dot{S}&=&-{\rm Tr}(\dot{\hat{\rho}}_{FF}\ln \hat{\rho}_{FF})\nonumber\\
&=&-{\rm Tr}\Bigl( \Gamma(R)\hat{\rho}_{FF}(e^{\beta\hbar\omega}\hat{a}^+\hat{a}-\hat{a}\hat{a}^+)\ln \hat{\rho}_{FF}\Bigr)\nonumber\\
&=&2\beta^2\frac{\hbar^4}{m^2}\frac{e^{\beta\hbar\omega}}{(e^{\beta\hbar\omega}-1)^2}
\frac{\dot{R}}{R^5}.
\end{eqnarray}
This result means that $\dot{S}\ge 0$ when the confining potential of HO is becoming wider (: $\dot{R}\ge 0$), which is physically reasonable.

Next, combining $\dot{S}$ with
the entropy production rate of the reservoir $-\beta \dot{Q}$ with $\dot{Q}$ in Eq.(\ref{t-der heat}), we can evaluate 
the entropy production rate $\dot{\sigma}$ of 
the whole system \cite{Ali2}:
\begin{eqnarray}\label{t-der entropy}
\dot{\sigma}\equiv\dot{S}-\beta\dot{Q}=0.
\end{eqnarray}
The final result can be proved by using the equality
$\frac{\partial}{\partial t}\coth(\beta\hbar\omega/2)=\frac{4\beta\hbar^2}{m}\frac{\dot{R}}{R^3}\frac{e^{\beta\hbar\omega}}{(e^{\beta\hbar\omega}-1)^2}$.

Owing to the FF variant of DBC in Eq.(\ref{DBE-Coeff}), therefore, the entropy production rate of the whole system is kept zero during the time evolution, and a rapid temporal change of the whole system is reversible.  

The standard DBC in Eq.(\ref{KMS-0}) comes from the balance between the positive probabilities of emission ($\gamma_{\downarrow}(R)=\gamma_0 (1+n(\omega(R)))$) and absorption ($\gamma_{\uparrow}(R)=\gamma_0 n(\omega(R))$) of thermal bosons with the distribution $n(\omega(R))=\frac{1}{e^{\beta\hbar\omega(R)}-1}$. The FF variant of dissipative coefficients $\gamma_{\downarrow}^{FF}$ and $\gamma_{\uparrow}^{FF}$ should also be attributed to the positive probabilities of transition between neighbouring adiabatic states. Under this remark, we can show the existence of the speed limit of the FF driving.
$\gamma_{\downarrow}^{FF}$ and $\gamma_{\uparrow}^{FF}$ were obtained in Eq. (\ref{FF-dissi coeffs}) with $g_{\downarrow}$ and $g_{\uparrow}$ in Eq.(\ref{explicit g}). Taking the mean of the ratios $\frac{\gamma_{\alpha}^{FF}}{\gamma_{\alpha}}$, we have
\begin{eqnarray}\label{scaled prob}
\frac{\gamma^{FF}}{\gamma}\equiv \frac{1}{2}\sum_{\alpha=\downarrow,\uparrow}\frac{\gamma_{\alpha}^{FF}}{\gamma_{\alpha}}=1+\Delta(R, \dot{R}).
\end{eqnarray} 

Below, we choose a nontrivial value $c=0$ among the arbitrary constant $c$ in Eq.(\ref{explicit g}) and introduce $\tau_{relax}\equiv  \sum_{\alpha=\downarrow,\uparrow}\frac{1}{\gamma_{\alpha}}$ as a relaxation time towards the Gibbs state prior to the FF periodic driving. Then $\Delta(R, \dot{R})$ is written as
\begin{eqnarray}\label{Delta}
\Delta(R, \dot{R}) &=&\beta\frac{\hbar^2}{m}\dot{R}R^{-3}
 \frac{e^{\beta\hbar\omega(R)}}{(e^{\beta\hbar\omega(R)}-1)^2} \tau_{relax}\nonumber\\
& =&\frac{1}{2}\tau_{relax}\frac{d}{dt}\Bigl(\frac{1}{e^{\beta\hbar\omega(R)}-1}\Bigr),
\end{eqnarray}
 where the final expression is available from Eq.(\ref{width}).

Because of the positivity of $\frac{\gamma^{FF}}{\gamma}$ in Eq.(\ref{scaled prob}), we must impose $1+\Delta(R, \dot{R})\ge 0$.
In case that $R$ and $\dot{R}$ are trigonometric functions of $t$ as in Eqs. (\ref{vt}) and (\ref{lambda2}), 
$\Delta(R, \dot{R})$ oscillates between $\Delta_{max}(>0)$ and $\Delta_{min}(<0)$. Therefore the equation
\begin{eqnarray}\label{Delta-min}
|\Delta_{min}|=1
\end{eqnarray} 
determines the speed limit $\bar{v}_C$ of the FF periodic driving.

Introducing the scaled time $x$ together with the dimensionless inverse temperature $\tilde{\beta}$ and speed parameter $\eta$,
\begin{eqnarray}\label{dimless param}
x&=&\frac{t}{T_{FF}},\nonumber\\
\tilde{\beta}&=&\frac{\beta \hbar^2}{m R_0^2},\nonumber\\
\eta&=&\frac{\bar{v}\tau_{relax}}{R_0},
\end{eqnarray} 
$\Delta(R, \dot{R})$ in Eq.(\ref{Delta}) is rewritten as
\begin{eqnarray}\label{Delta-x-rep}
\Delta(x)=\frac{1}{2}\eta\frac{d}{dx}\Bigl(\frac{1}{\exp[\tilde{\beta}(1+\frac{1}{2\pi}(1-\cos(2\pi x)))^{-2}]-1}\Bigl),\nonumber\\
\end{eqnarray} 
where the prescription in Eq.(\ref{mean veloc}) is used. The minimum of $\Delta(x)$ is given at $\tilde{\beta}$-dependent extreme value $x_0$ as
\begin{eqnarray}\label{Delta-min-rep}
\Delta_{min}=\Delta(x_0)\equiv-\eta\tilde{\beta}Y_0(\tilde{\beta}).
\end{eqnarray} 

\begin{figure}[!h]
\begin{center}
\includegraphics[width=3.5in]{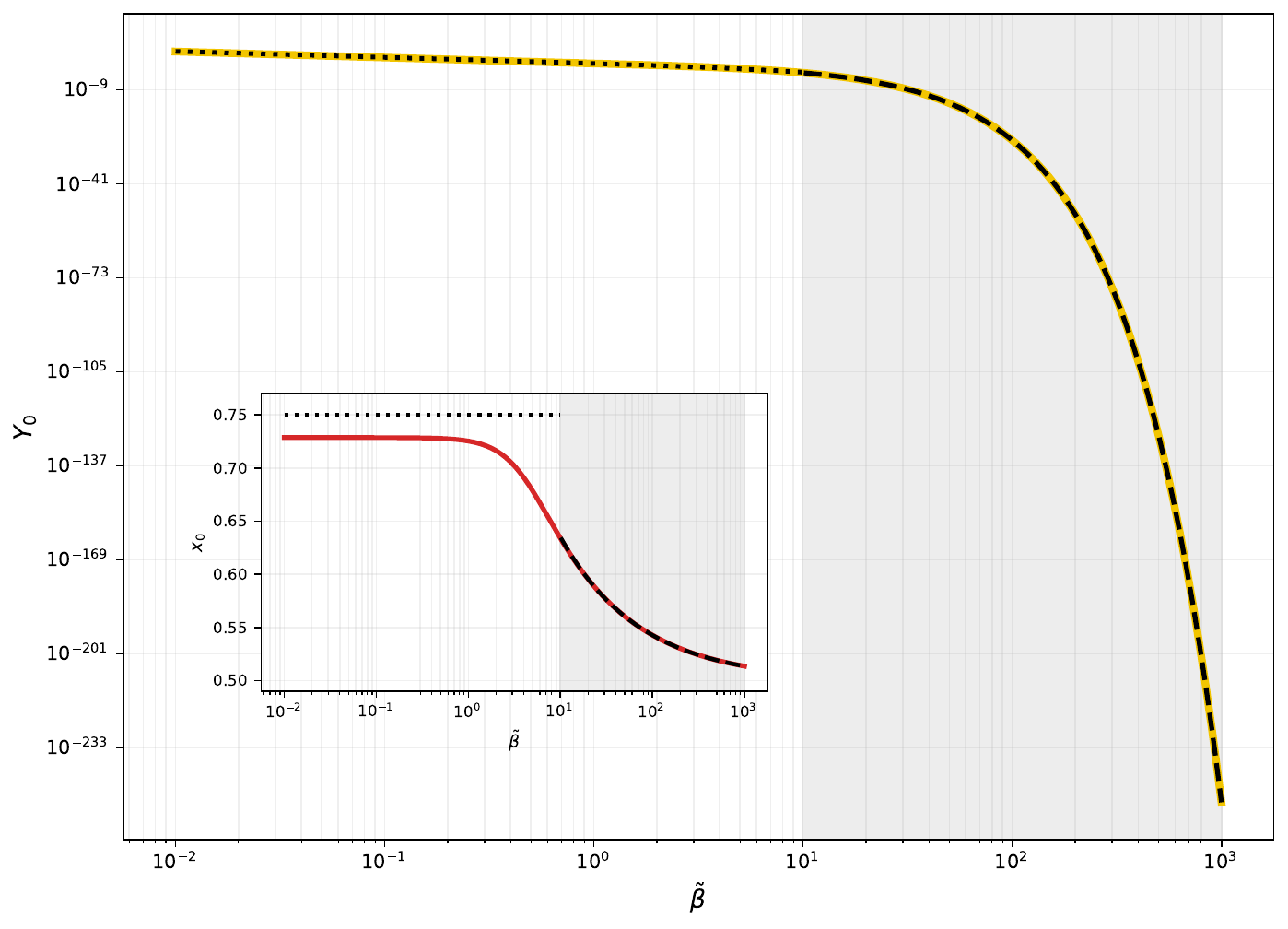}
\caption{$Y_0\equiv \Delta(x_0)/(-\eta \tilde{\beta})$ versus $\tilde{\beta}$.
Colored solid curve presents numerical results. Dotted and broken lines show the best-fitting functions in Eq.(\ref{Delta-min-y})
in the ranges $10^{-2}\leq \tilde{\beta} <10$ and $10\leq \tilde{\beta} \leq 10^3$, respectively.
Inset: numerical results on $x_0$ versus $\tilde{\beta}$. The range $\tilde{\beta}=10^{-2}\sim10^{3}$ corresponds to temperature 
$T=10^{-3}K\sim10^2 K$ with use of proton mass $m\sim10^{-27} kg$ and confinement length $R_0\sim1nm$.} 
\label{Iwan asympt}
\end{center}
\end{figure}
Figure \ref{Iwan asympt} presents numerical results of $x_0(\tilde{\beta})$ and $Y_0(\tilde{\beta})$
and the corresponding best-fitting functions $Y_0(\tilde{\beta})=a\tilde{\beta}^{-n}\exp(-b\tilde{\beta})$
with
\begin{eqnarray}\label{Delta-min-y}
Y_0(\tilde{\beta})&=&1.3652\tilde{\beta}^{-2}\exp(-0.2547\tilde{\beta})    \qquad     10^{-2}\leq \tilde{\beta} <10,\nonumber\\
Y_0(\tilde{\beta})&=&0.7102\tilde{\beta}^{-1/2}\exp(-0.5754\tilde{\beta})    \qquad    10\leq \tilde{\beta} \leq 10^{3}.\nonumber\\
\end{eqnarray}

The combination of Eqs.(\ref{Delta-min}), (\ref{Delta-min-rep}) and (\ref{Delta-min-y}) determines the upper limit of the speed $\eta_c$ which is accessible by the FF periodic driving:
\begin{eqnarray}\label{speed uplimit}
\eta_c&=&0.7325\tilde{\beta}\exp(0.2547\tilde{\beta})    \qquad     10^{-2}\leq \tilde{\beta} <10,\nonumber\\
\eta_c&=&1.4081\tilde{\beta}^{-1/2}\exp(0.5754\tilde{\beta})    \qquad    10\leq \tilde{\beta} \leq 10^{3}.\nonumber\\
\end{eqnarray} 
$\eta_c$ gives the upper velocity limit $\bar{v}_c=\frac{R_0}{\tau_{relax}}\eta_c $ from Eq.(\ref{dimless param}) and the lower limit of the periodicity $T^c_{FF}=\frac{\tau_{relax}}{\eta_c}$ with use of Eqs.(\ref{mean veloc}) and (\ref{dimless param}).
$\eta_c$ increases as the temperature is lowered. In the asymptotic region  $\tilde{\beta}\ll 1$,  $\eta_c \sim \tilde{\beta}$.
This tendency remains unchanged for general choice of constant $c$ so long as $c > -e^{-\beta\hbar\omega}$.
 
The same issue as in this Section can be available for a dissipative quantum spin system in Section \ref {AOQD2}.

\section{Summary and discussions}\label{conclude}
To create a non-equilibrium Gibbs state, we have applied the scheme of the fast forward-and-backward (abbreviated as FF for convenience) to driven dissipative quantum  systems described by GKSL equation. Taking a harmonic oscillator (HO) under a rapid time-periodic change of its confinement parameter and  in contact with the heat reservoir, we revealed a protocol which guarantees the non-equilibrium Gibbs state.  

Analysis begins with the Gibbs state under the fixed confinement parameter, prepared in advance under the standard detailed balance condition (DBC) or KMS condition among a pair of dissipative coefficients, and then we let the parameter time-dependent.  Firstly, assuming the quasi-adiabatic modulation of the parameter and  introducing  the Gibbs state which depends on time implicitly through the parameter, we solve the inverse problem to obtain the non-adiabatic corrections to Hamiltonian and to the dissipator. Secondly, to speed up the quasi-adiabatic dynamics and to recover the dynamics on ordinary time scale,  we apply the FF time-periodic rescaling, obtaining the FF variant of DBC  in Eq.(\ref{DBE-Coeff}) as well as the FF Hamiltonian. By taking the FF time ($T_{FF}$) as a period of the periodic driving, we obtain the time-periodic Gibbs state in the dissipative environment under the FF protocol. To indicate the universality of the FF scheme,  we applied it to another prototype case, i.e., a driven dissipative quantum spin system and obtained the same FF variant of DBC in Eq.(\ref{FF DBC}) as in the harmonic-oscillator case.
In the FF approach so far, our target has been the adiabatic eigenstate of Hamiltonian in non-dissipative unitary dynamics. By contrast, this article is concerned with adiabatic modulation of the fixed-point solution of the GKSL equation and  its FF variant under the time-periodic driving field. 

To see a role of the FF variant of DBC,  we evaluated the power, heat current and entropy production rate  of the HO system under the FF driving in contact with the heat reservoir at constant temperature, and confirmed the energy conservation of power and heat current of the system and reversibility of whole system including a reservoir.
The requirement of positivity of the FF dissipative coefficients imposes a constraint on the speed of the FF periodic driving.  Analyzing them, we find that the upper speed limit  of the driving
increases as temperature is lowered and that at low temperature this limit is proportional to $\tilde{\beta}^{-1/2}\exp(a\tilde{\beta})$ with $\tilde{\beta}$ the scaled inverse temperature and $a$ the positive constant of $O(1)$.

Since the extra Lindblad dissipators are shown to depend on the driving protocol, we must proceed to a microscopic derivation of the FF dissipative coefficients, so that such coefficients are compatible with a thermal bath or a Born–Markov derivation. Our future challenge will be directed towards these themes.

\section*{Acknowledgment}
K. N. is grateful to K. Takahashi and S. Abe for fruitful discussions in the early stage of the work. Our  special thanks are devoted to T. Monnai who read the original version of this manuscript and gave us critical comments. I.S. thanks to RIIM grant   contract number 2529/UN30.15/KS/2026 from Indonesian Agency of Research and Innovation 2026.

\let\doi\relax

\appendix  
\section{Equivalence and difference of dissipators between STA and FF approaches}\label{apdB}
Applying the STA scheme, Alipour et al \cite{Campo1} presented a model of driven system in the presence of balanced gain and loss. However, they neither included a primary dissipator satisfying the KMS condition nor addressed an improved version of detailed balance condition. Besides, they paid no attention to a rapid periodic driving. Nevertheless, it is worthy to examine here the similarity and difference between the STA and FF approaches in the context of the quantum master equation. 

Their GKSL equation, which employs the Hermitian ($\hat{x}$) operator(instead of boson operators) as the jump operator and works for the unitary-transformed version $\tilde{\rho}$ of the Gibbs state $\rho(t)=e^{-\beta H_0}/{\rm Tr}(e^{-\beta H_0})$, was given in the non-rotating frame as follows (See Eqs.(43) and (44) in \cite{Campo1}):
\begin{eqnarray}\label{STA mast eq}
\frac{\partial \tilde{\rho}}{\partial t}=-\frac{i}{\hbar}\big[\frac{\hat{p}^2}{2m}+\frac{1}{2}m\tilde{\omega}^2_{CD}\hat{x}^2, \tilde{\rho}\big]
-\gamma[\hat{x},[\hat{x},\tilde{\rho}]],
\end{eqnarray}
where 
\begin{eqnarray}\label{STA omg-coeff}
\tilde{\omega}^2_{CD}=\omega^2-\delta^2 -\dot{\delta},\nonumber\\
\gamma=\frac{m\omega}{\hbar}\frac{\dot{u}}{(1-u)^2},
\end{eqnarray}
with $\delta=-\frac{\dot{\omega}}{2\omega}$ and $u=e^{-\beta\hbar\omega}$. 
To avoid the overlap of notations between the present paper and theirs, the symbol $\delta$ is here used instead of $\alpha$ in \cite{Campo1}.

To compare Eqs.(\ref{STA mast eq}) and (\ref{STA omg-coeff}) with Eqs.(\ref{FF-GKSL})-(\ref{DBE-Coeff}) in our  text, we must construct an  inversely unitary-transformed version of Eq.(\ref{STA mast eq})  by using $U=e^{i\frac{m}{2\hbar}\delta\hat{x}^2}$.
Introducing the STA density operator
$\rho_{STA}\equiv U^{+}\tilde{\rho}U$, and noting the equalities
$U^+\hat{p}^2U=\hat{p}^2+m\delta\{\hat{x},\hat{p}\}+m^2\delta^2\hat{x}^2$ and
$U^+\hat{x}U=\hat{x}$, we can obtain the re-constructed master equation for $\rho_{STA}$:
\begin{eqnarray}\label{rho STA}
\frac{\partial \rho_{STA}}{\partial t}=-\frac{i}{\hbar}\big[H_{STA}, \rho_{STA}\big]
-\gamma[\hat{x},[\hat{x},\rho_{STA}]],
\end{eqnarray}
where
\begin{eqnarray}\label{H-STA}
H_{STA}=\frac{\hat{p}^2}{2m}+\frac{m}{2}\omega^2_{CD}\hat{x}^2+\frac{\delta}{2}\{\hat{x},\hat{p}\}
\end{eqnarray}
with 
$\omega^2_{CD}=\tilde{\omega}^2_{CD}+\delta^2+\dot{\delta}=\omega^2$.
$H_{STA}$ is equal to $H_{FF}$ in Eq.(\ref{FF-Haml}). See also Refs. \cite{STA-0} and \cite{NMG}.
In the STA scheme, however, the protocol function $\omega(t)$ under the boundary condition at the initial and final time must satisfy a global time-dependent constraint like the Ermakov equation. By contrast, the FF scheme can employ any one of arbitrary smooth and continuous functions of $\omega(t)$ satisfying only the boundary condition, as noted above Eq.(\ref{alp-scal}). 

As for the dissipator, we must rewrite the second term on the right-hand side of Eq.(\ref{rho STA}) in terms of creation ($\hat{a}^+$) and annihilation ($\hat{a}$) operators available from the inverse of the relation just below Eq.(\ref{diagonal}),
and suppress the contribution of double-bosons creation and annihilation processes. Then the dissipator in Eq.(\ref{rho STA}) is approximated as
\begin{eqnarray}\label{dissip1-STA}
-\gamma[\hat{x},[\hat{x},\rho_{STA}]]\simeq \sum_{\alpha=\uparrow,\downarrow}\gamma_{\alpha}^{STA}\times \nonumber\\
\Bigl(\hat{L}_\alpha \rho_{STA}\hat{L}^+_\alpha-\frac{1}{2}\{\hat{L}^+_\alpha \hat{L}_\alpha,\rho_{STA}\}\Bigr),
\end{eqnarray}
where
\begin{eqnarray}\label{gamma-STA}
\gamma_{\uparrow}^{STA}&=&\gamma_{\downarrow}^{STA}=\gamma\frac{\hbar}{m\omega}=\frac{\dot{u}}{(1-u)^2}\nonumber\\
&=&2\beta\frac{\hbar^2}{m}\dot{R}R^{-3}e^{\beta\hbar\omega}/(e^{\beta\hbar\omega}-1)^2.
\end{eqnarray}
Equation (\ref{dissip1-STA}) corresponds to the additional part of the dissipator 
$\dot{R}d[\hat{\rho}_{FF}]$ in Eq.(\ref{FF-Dissip}), while Eq.(\ref{gamma-STA}) is equivalent to the common factor of the FF variant of extra dissipative coefficients 
on the right-hand side of Eq.(\ref{explicit g})
multiplied with $\dot{R}$.
The dissipative coefficients must satisfies
$\gamma_{\uparrow}^{STA}=\gamma_{\downarrow}^{STA}$ in the STA scheme, whereas there is no such constraint on
the additional part of the dissipator 
$\dot{R}d[\hat{\rho}_{FF}]$ in Eq.(\ref{FF-Dissip}) in our text. Namely, the constant $c$ in Eq.(\ref{explicit g}) is arbitrary.

\end{document}